A Qualitative Exploratory Study into Vulnerability Chaining Blindness Terminology and Viability

Nikki Robinson, Capitol Technology University, nerobinson@captechu.edu

Precis: The researcher conducted a focus group of 6 cybersecurity and IT professionals to investigate the phenomenon of professionals inability to view and remediate multiple vulnerabilities at one time. The researcher introduces the new terminology *vulenrability chaining blindness* to describe this phenomenon

Author Biography: Nikki Robinson holds a DSc in Cybersecurity, as well as a PhD in Human Factors from Capitol Technology University. She is a Security Architect with IBM as well as an Adjunct Professor for Capitol Technology University. She hold several industry certifications, including CISSP and CEH, and specializes in vulnerability management, digital forensics, incident response, and integrating human factors with cybersecurity. Her research is primarily focused on understanding how cybersecurity professionals understand and mature vulnerability management programs, as well as how human factors principles can improve current cybersecurity practices.

*Abstract* - To tie together the concepts of linkage blindness and the inability to link vulnerabilities together in a Vulnerability Management Program (VMP), the researcher postulated new terminology. The terminology of vulnerability chaining blindness is proposed to understand the underlying issues behind vulnerability management and vulnerabilities that can be used in combination. The general problem is that IT and cybersecurity professionals have a difficult time identifying chained vulnerabilities due to the complexity of vulnerability prioritization and remediation (Abomhara & Køien, 2015; Felmetsger et al., 2010). The specific problem is the inability to link and view multiple vulnerabilities in combination based on limited expertise and awareness of vulnerability chaining (Tang et al., 2017). The population of this study was limited to one focus group, within the IT and Security fields, within the United States. The sample size consisted of one focus group comprised of 8-10 IT and cybersecurity professionals. The research questions focused on if participants were aware of linkage blindness or vulnerability chaining, as well as if vulnerability chaining blindness would be applicable to describe the phenomenon. Several themes emerged through top-level, eclectic, and second-level coding data analysis. These themes included complexity in cybersecurity programs, new concepts in vulnerability management, as well as fear of the unknown and where security meets technology.

*Keywords: linkage blindness, vulnerability chaining, vulnerability chaining blindness, vulnerability management*

## Introduction

The field of Human Factors Engineering (HFE) research is focused on engineering in the analysis and design of human-technological systems (Phillips et al., 2006). The primary motivation for this science is to improve or optimize systems performance based on human's limitations or capabilities while using that system (Phillips et al., 2006). As the complexity of machines and systems grows, our demands as users and operators increase as well (Guastello, 2014). There are both challenges and potential opportunities between humans and computer systems (Guastello, 2014).

Given these potential challenges, HFE research combines the efforts of researchers, engineers, and operators. Researchers can improve knowledge in the field of HFE by examining the implications between the human operators and the systems required for business or tasking (Guastello, 2014). HFE research was expanded to include everything from applied cognitive psychology, technology, human computer interaction, and aviation psychology (USD, 2020). Since this field is ever-expanding and on the forefront of research for computer systems, this field is also applicable to Information Technology (IT) and cybersecurity practices.

The nuanced area of vulnerability management within HFE research will be the focus of this dissertation. While the research in Cybersecurity covers a wide range of topics (Akhunzada et al., 2015; Moşteanu, 2020; Pan & Yang, 2018; 2015; Syed, 2020), comprehension of vulnerabilities and vulnerability management are of particular interest. Rico, Engstrom and Host (2019) noted the need to create a taxonomy to improve industry-academia communication around vulnerability management. This revelation led the researcher to believe taxonomies were missing in other parts of vulnerability management and remediation.

## Background

Linkage blindness is a term used in law enforcement to describe the inability of detectives to communicate information that may connect similar crimes (Sullivan, 2009). The problem was researched by Egger (1984) to define serial murder and reduce linkage blindness between law enforcement entities. Egger noted the difficulty for law enforcement to solve serial murders or determine serial murders are occurring based on different geographic locations and lack of communication between law enforcement agencies. The terminology *linkage blindness* was developed in 1984 to represent serial murder investigations and the inability to solve them (Hickey, 2003). This taxonomy was created specifically for criminal investigations to identify issues and solve crimes, but does this phenomenon occur in other fields of study?

Baechler et al. (2012) noted that linkage blindness applies to the creation and use of false identify documents used by organized crime and terrorist organizations. The framework proposed by Baechler et al. would detect links and patterns where false identify documents are used in forensic intelligence processes. The use of linkage blindness in this study displayed the need in other forensics sciences to link criminal behavior and identify associated patterns in criminal organizations. The inability to link crimes together, whether in serial murders or false identify documentation, showed that this phenomenon required further research.

The concept of individuals in a field unable to link data together, which led to a more serious concern, was applicable to other industries or types of problems. The researcher considered expanding this taxonomic concept from law enforcement to the field of IT and cybersecurity. Cybersecurity threats and malicious actors increased in sophistication at an alarming rate (Tounsi & Rais, 2018). Network engineers, security engineers, and IT professionals must increase security defenses to adapt to these threats (Tounsi & Rais, 2018). One area of cybersecurity that required more research both academically and in the private sector is vulnerability management (Evans et al., 2016).

## Problem / Purpose

**Problem Statements**

The general problem is that IT and cybersecurity professionals have a difficult time identifying chained vulnerabilities due to the complexity of vulnerability prioritization and remediation (Abomhara & Køien, 2015; Felmetsger et al., 2010). Absent from the literature is extensive research into vulnerability chaining as it related to IT and Security professional's comprehension and ability to remediate those vulnerabilities effectively. There was an incredible amount of current and ongoing research on improving vulnerability scoring (Allodi & Massacci, 2017; Ganin et al., 2017; Mantha et al., 2020; Roldán-Molina et al., 2017) to aid vulnerability management programs. However, concerns around vulnerability chaining were still a more researched topic in the private and public sectors instead of in academic studies (FIRST, 2020).

The specific problem is the inability to link and view multiple vulnerabilities in combination based on limited expertise and awareness of vulnerability chaining (Tang et al., 2017). This area of study has a narrow focus in cybersecurity, but was seen in other fields like law enforcement, which meant that there is room for additional research in new fields. As the current state of literature was missing an exploration into vulnerability chaining and the need for a new taxonomy, the researcher postulated this could have a significant impact in both the IT and cybersecurity fields. The implications of a new terminology to describe *vulnerability chaining blindness* could increase the awareness of vulnerability chaining and aid in the remediation of vulnerabilities.

**Purpose**

The purpose of this study was to conduct a phenomenological inquiry into the inability to link vulnerabilities when defending a network. More specifically, how do IT administrators and security analysts view vulnerabilities and prioritize remediation. Based on current knowledge of the field and professional experience, the researcher saw both IT and cybersecurity teams review and remediate vulnerabilities in a singular manner. This first-hand knowledge allowed the researcher to investigate the possibility of this phenomenon and complexities around vulnerability management.

CVSS defined vulnerability chaining as the scoring of multiple vulnerabilities when used in combination to exploit during a single cyberattack (FIRST, 2020). This chain of vulnerabilities was used to compromise hosts, applications, or an entire network. While not exclusively about vulnerability chaining, a well-known TTP (Tactics, Techniques, and Procedures) model is the Cyber Kill Chain developed by Lockheed Martin

(2021). The Cyber Kill Chain identified the methods used by attackers to aid in network defender's awareness of vulnerability and exploitation use in a cyberattack (Lockheed Martin). CVSS, the Cyber Kill Chain, and even the MITRE ATT&CK framework all identify vulnerability chaining as a common technique of attackers.

Linkage blindness has been identified and used in law enforcement, so the main question was, does this concept apply in other fields? Of specific interest, does linkage blindness apply to vulnerability identification in the cybersecurity field? These implications could lead to broader awareness of this problem, and address this issue with a new term, vulnerability chaining blindness. If this terminology was deemed relevant and applicable, it could allow IT and cybersecurity professionals to identify vulnerabilities in combination. This identification of multiple vulnerabilities leading to one or more exploits could potentially lead to the ability to stop more complex attacks from malicious actors.

The study addressed the missing research and lack of taxonomy to describe the issues surrounding vulnerability identification and remediation. The purpose of this exploratory design was to use qualitative methods to collect information from both IT and security professionals. Through this data collection, participants were presented with the idea of *vulnerability chaining blindness* and to determine where this issue exists within the organization. This study explored the concepts of linkage blindness and the application in vulnerability remediation within the cybersecurity field. Using a focus group qualitative design, both IT and cybersecurity professionals were given a set of questions surrounding this phenomenon.

## Research Method

Quantitative research and mixed methods were not appropriate for this study. Quantitative research is focused on gathering numerical data and conducting a statistical analysis. This research aimed to define a phenomenon and explore whether new terminology could be used to describe the inability to view and remediate multiple vulnerabilities simultaneously. Mixed methods research combines both qualitative and quantitative designs to gather numerical data and then use open-ended questions to participants to determine how or why they answered in a particular way. As there was no prior research or literature about *vulnerability chaining blindness*, there was no need to collect quantitative data currently. This led the researcher to determine a qualitative research design is the most appropriate.

**Research Questions**

The research questions are an essential component of the research design. These questions help to explore the general and specific problem statements and identify in more detail where these problems exist (Creswell & Creswell, 2018). Through the questions stated below, the researcher investigated the topic of linkage blindness as it related to vulnerability chaining. Of special interest was not only if linkage blindness is applicable, but do IT and cybersecurity professionals agree on its viability in the field? The research questions were used to map trends and patterns back to the focus group responses.

*RQ1*: Does the concept of linkage blindness apply to the inability of security professionals to view multiple vulnerabilities at a time?

*RQ2*: Does the terminology of *vulnerability chaining blindness* address the concerns of vulnerability chaining and vulnerability remediation?

*RQ3*: Is the concept of *vulnerability chaining blindness* an appropriate way of addressing vulnerability chaining identification concerns?

*RQ4*: Are IT and cybersecurity professionals aware of, and able to articulate, the concept of vulnerability chaining used by hackers?

*RQ5*: Can IT and cybersecurity professionals use *vulnerability chaining blindness* to improve vulnerability remediation practices?

*RQ6*: Are IT and cybersecurity professionals in agreement on the concept of *vulnerability chaining blindness*?

**Population**

The population for this study was based on IT and cybersecurity professionals in the United States (US). The list of relevant job titles for IT professionals includes Systems Administrator, Network Engineer, IT Operations, or Helpdesk (Tier I/II), or Chief Information Officer (CIO). Relevant job titles for cybersecurity professionals included Security Operations Center (SOC) analyst, Security Engineer or Architect, Information Systems Security Officer (ISSO), or Chief Information Security Officer (CISO). These job titles and descriptions encompassed a wide range of skills to take part in this focus group. The intention for having both IT and cybersecurity professionals in the focus group was to determine if both fields agreed that the phenomenon of *vulnerability chaining blindness* occurred in their respective fields.

**Sample**

A sample was taken from the researchers' LinkedIn community which comprised of over 700 IT, cybersecurity, and executive management. Participants had to be over 18 and at least a year of experience within either the IT or cybersecurity industry. This allowed for a larger pool of participants because hashtags were used to encourage even more LinkedIn members to participate. Once individuals expressed interest in the study, the researcher compiled a list of possible participants in a spreadsheet. The researcher used a Random Number Generator (RNG) to select 10 individuals to take part in the focus group. This procedure removed the possibility of bias during the participant selection phase.

**Data Analysis**

Several methods of qualitative coding analysis were used to create a holistic view of the themes and patterns throughout the focus group responses. Values coding as used as a top-level coding method, next was eclectic coding, and finally axial coding was used to identify overall patterns. Each level of coding gave new insight and organized the codes from the participant responses.

**Values Coding**

Overall, Participant 1 was not as responsive as other participants of this research but provided thoughtful insight with questions and concerns. Participant 1 did place value on how helpful this research would be to IT and cybersecurity professionals, specifically with the quote, "Is it developing a common vernacular across professionals, is it updating things, or taking this in a different direction…". The coded values from this participant and statement were *concern* and *useful*. This response was in direct relation to if *vulnerability chaining blindness* would be used to explain the phenomenon of vulnerability chaining.

Participant 2 was one of the most interactive participants in the focus group. This participant placed value on focusing on cybersecurity holistically, and not focusing on individual vulnerabilities. Participant 2 placed value on *layered defense* and ensuring that individuals were not focused on just vulnerability chains but more about *unknown unknowns*. This terminology is used in cybersecurity to describe unknown threats of an unknown type, which are impossible to defend against. However, when it came to understanding vulnerability chaining blindness, there was value placed on understanding low and medium vulnerabilities as a consideration. This value is exemplified by the quote, "if somebody can take several low criticality vulnerabilities and ignore them and if they can string them together and form an attack using that I think that there's likely, a lot of organizations that are vulnerable because of that."

Participant 4 was also very vocal within the focus group and placed value on *agility* and *speed* when applying patches or resolving vulnerabilities quickly. This response came from the question of resolving vulnerabilities one at a time or in combination. Another question where the values of *agility* and *speed* came in were regarding automating repetitive tasks. This direct quote explains the requirement from a software development mindset, "I would think it would because so from my space. You know if you are monitoring and network. You know, one of the things that. You know that we focus on is signature ability and being able to identify something repeatedly".

Participant 6 was also continually active in the focus group and provided insight from a threat intelligence perspective. Participant 6 placed value on understanding the *kill chain* and how *threat intelligence* helps to direct cybersecurity and vulnerability management programs. The importance of understanding these concepts was highlighted in this direct quote, "So, if you look at it from the view of the threat actor and. The attack kill chain right so there is always a when you are doing the postmortem right and you are coming back and hunting down the threat actors inside of your environment, you are trying to trace them steps back through the kill chain, you know, some people will tie that to the vulnerability."

Participant 7 placed more value on the concept of *slow and steady* when it came to applying patches, as from the IT perspective it is easier to troubleshoot issues when patches are applied one at a time. This was in direct opposition from the software development or cybersecurity perspective and offered additional insight into why vulnerabilities may be remediated one at a time. There was additional value placed on *troubleshooting* and how this concept plays into patching systems, but also how functionality concerns from users may lead to delayed patching.

**Attitudes Coding**

While Participant 1 had fewer responses overall, one clear attitude came across in each response. There was a general positive attitude towards including law and policy terminology into cybersecurity practices. Based on their graduate work, they felt that there was a connection between how law enforcement terminology, such as *chain of evidence*, could be used in the way cybersecurity incidents are handled. This direct quote highlights the general positive attitude towards including legal and cybersecurity terminology, "another example would be how the FBI basically funds local labs at local, law enforcement, give them the ability to understand how to be able to collect digital evidence, without compromising it, so I would say yes, both inside and outside of IT and cyber security."

Participant 2 had an attitude of *fear* when it came to focusing solely on vulnerability chaining examples in an organization. The response was an expression of concern around missing a bigger picture, which would lead to 'being afraid' of missing critical vulnerabilities to focus on lower or medium vulnerabilities. This direct quote from Participant 2 uses the direct terminology related to fear, "I would be afraid, if you're focusing on that you're going to miss the bigger picture like really we should all be building a layer to Defense model." There was also a fear related to the risk associated with not paying attention to *vulnerability chaining blindness*.

Participant 4 had a general attitude of being *excited* and *interested* in *vulnerability chaining blindness* and its widespread applicability. They mentioned how the terminology could be broadened beyond IT and cybersecurity to understanding vulnerabilities from a more technical perspective. Participant 4 noted that, "[The] technical vulnerability that's in there, so I think it has broader applications than just any other [vulnerabilities] applied together." This sentiment was echoed throughout the focus group questions, but the last quote directly reflects the participants believe in the terminology's use outside of technical fields.

Participant 6 however noted that organizations may use *finger pointing*, or *blaming*, in different parts of the business when an incident occurs. This was an interesting attitude that was expressed by both Participant 2 and Participant 6. This quote from Participant 6 is like the blaming experienced by Participant 2, "With regards to pointing fingers right different departments and different parts of the organization. Whenever there is one of these big major catastrophic vulnerabilities that gets exploited." Because this attitude was shared between two participants, this is an item that will be explored further in eclectic and second-level coding.

Participant 7 also expressed *fear* when it came to applying multiple patches or resolving multiple vulnerabilities at one time. Participant 2 also shared this view of concern and *fear* when it came to the same question, although Participant 2 was more concerned about the larger picture of vulnerability management. Participant 7 noted that, "it can be many times worse if something breaks than just changing permissions or changing one or two active directory objects." As two participants expressed an attitude of fear on similar topics in the focus group, these codes and patterns will also be explored further in the later coding sections.

**Belief Coding**

  Participant 1 had a strong belief in the concepts of *law* and *criminal justice* as it related to cybersecurity principles presented in the focus group. There were several quotes that explicitly were concerned with the inability to understand terminology and its impacts when prosecuting criminals and how that relates to the concept of *vulnerability chaining blindness*. Specifically, this quote highlights the belief in combing law and terminology, "Even update law we think about prosecution against cyber criminals there isn't anything in the law today." This strong belief was rooted in their professional experience and drawn from the problems seen in the lack of associated cybersecurity law terminology.

  Participant 2 stressed a belief in *focusing on owners and operators* when it came to resolving vulnerabilities and patching systems. This direct quote from Participant 2 showed changing priorities in IT Operations and Cybersecurity, "You know so then changed our priority to focus on the owners and operators of those two systems and figure, if we can pull them in line then it will be easier to you know chip away at the other 25%." However, this was not the only belief that Participant 2 expressed, they also mentioned the concept of *big picture* in terms of vulnerability management.

  Participant 4 also held a strong belief in the *big picture* of cybersecurity programs and not losing sight on singular vulnerabilities. Participant 4 noted that, "…you know if you focus too much on one, you're going to miss everything else." Since this belief was echoed by Participant 2 as well, this will be further explored in the eclectic and second-level coding sections. Another belief that stood out was that *full compliance is impossible* within a system. The quote from Participant 4 that detailed this belief was, "I never found one that I could actually still run software that we needed to run in 100% compliant system."

  Participant 6 expressed a belief in the *complexity of vulnerabilities* when it comes to how APT groups and malicious actors use vulnerability chaining attacks. This belief was outlined in several quotes, but specifically relating to how vulnerability chaining blindness could be applied. Participant 6 noted that, "they're going to use a variety of different tools until you find one that works inside of that you know he through each step of the framework." This quotation also highlights the belief in *persistence of attackers* when using multiple combinations of vulnerabilities in a cyberattack.

Participant 7 had a unique belief system structured on the *perception of users* when it came to blaming a specific application when any problem occurred. Participant 7 noted that, "After we would migrate from a site, suddenly, every connectivity issue was an active directory issue that the network, people were convinced that…" The response was in direct response to the applicability of linkage blindness concepts in cybersecurity programs. This led the researcher to believe there may be more within Participant 7's responses that should be evaluated further as more codes are defined.

*TABLE 1.*

| Top-Level Coding | Values, Attitudes, Beliefs Coding | | |
|---|---|---|---|
| | *Value* | *Attitude* | *Belief* |
| Participant 1 | Useful | Positivity | Law, Criminal Justice |
| Participant 2 | Layered Defense | Fear, Risk | Focusing on Owners and Operators, Big Picture |
| Participant 4 | Agility, Speed | Excitement, Interest | Big Picture, Full Compliance |

| Top-Level Coding | Values, Attitudes, Beliefs Coding | | |
| --- | --- | --- | --- |
| | *Value* | *Attitude* | *Belief* |
| | | | is Impossible |
| Participant 6 | Kill Chain, Threat Intelligence | Finger Pointing, Blame | Complexity of Vulnerabilities, Persistence of Attackers |
| Participant 7 | Slow and Steady, Troubleshooting | Fear, Risk | Perception of Users |

**Axial Coding**

The first axial code identified during this process was *fear of the unknown*. Participants expressed *concern*, *technical limitations*, *understanding*, and *fear*. When examining participant responses, it was clear that the concern was a fear of the unknown. Since the concepts of *linkage blindness* and *vulnerability chaining* were already new concepts, adding *vulnerability chaining blindness* compiled the fear. This was a concern the researcher had not considered, given the widespread use of the terminology *vulnerability chaining* in both NIST documentation and the CVSS calculator. But this led the researcher to determine that the concern coming from participants was not the use of *vulnerability chaining blindness*, but more that it was an untested concept.

Fig. 1. Fear of the Unknown

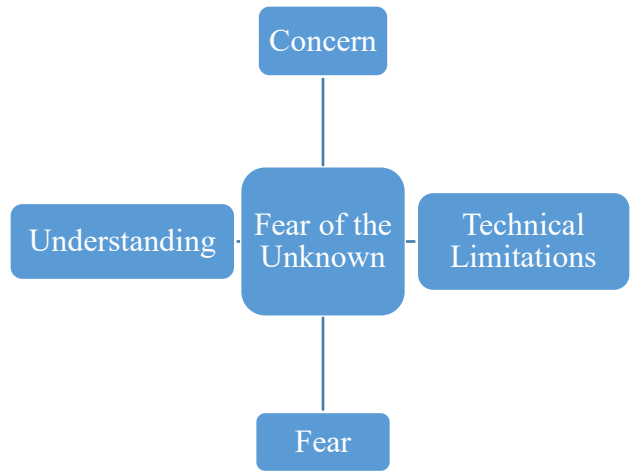

The second axial code from the initial and eclectic coding phases was *where security meets technology*. Participants expressed *technical concerns*, *threat intelligence*, *technology*, and *big picture* all as critical concerns in a cybersecurity program. Between IT and cybersecurity professionals, it was clear that security and technology, specifically operations, are an important combination. Both teams must work together to secure operational applications and infrastructure, to meet technology and cybersecurity objectives. A theme emerged that each team may perceive *vulnerability chaining blindness* differently, it was crucial for each of them to keep an open mind about how the terminology would be used.

Fig. 2. Where Security Meets Technology

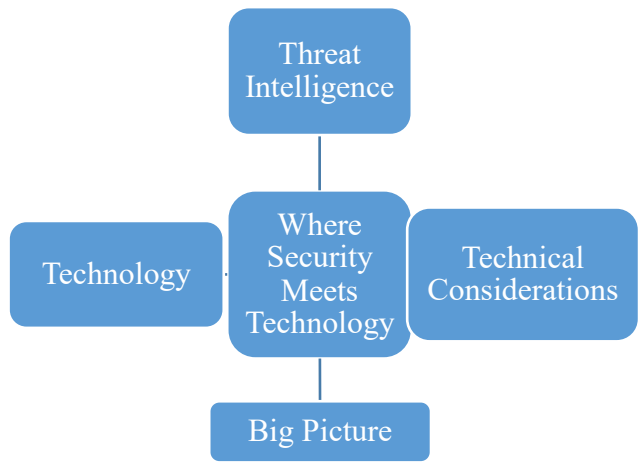

The third code that emerged during data analysis was *new concepts in vulnerability management*. When analyzing participants responses to multiple questions, the researcher noticed that several terms and phenomenon were novel. This code came from *prioritization of vulnerabilities*, *vulnerability chaining awareness*, *vulnerability chaining identification*, and *inability to link concepts*. Since both *linkage blindness* and *vulnerability chaining* were new terms for participants, it was a possibility that these new concepts could be applied to vulnerability management. Participants expressed positive interest in using *vulnerability chaining blindness* in their organizations with the understanding that this was a new concept and may take some time to implement.

*Fig. 3. New Concepts in Vuln Mgmt*

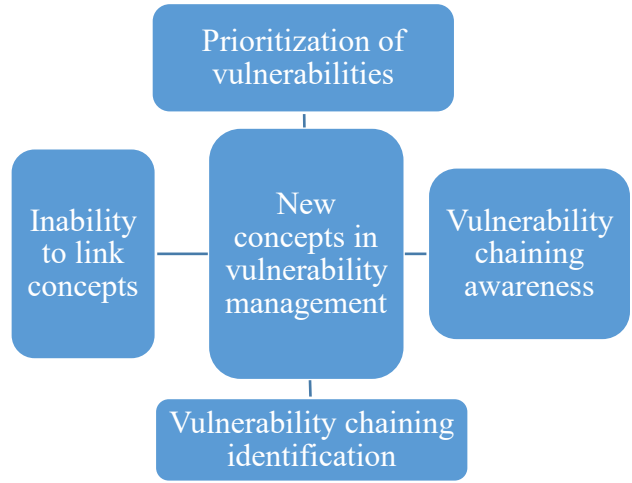

A final axial code identified during data analysis was the *complexity of cybersecurity programs*. The new terminology, concepts, and wide view of *vulnerability chaining* show that cybersecurity programs and understanding risk, are very complex. The code was developed using *perception of users*, *layered defense*, and *focus on system owners*. Cybersecurity practitioners must juggle technology, security controls, working on a layered defense, as well as how users and system owners perceive their systems. With the increase of cyberattacks, new types of malware and ransomware, not to mention new standards and technology all the time, cybersecurity is a difficult and ever-evolving field. It makes sense that participants mentioned a number of these items just within the context of *vulnerability chaining blindness*.

*Fig. 4. Complexity of Cybersecurity Programs*

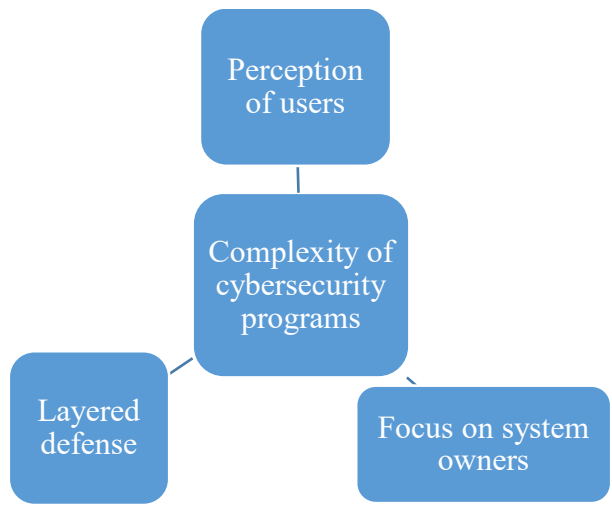

**Findings**

**Vulnerability Management**

    Vulnerability management was another fascinating component of the focus group. One participant began the focus group with describing how teams may place blame on IT operations groups based on how vulnerabilities are exploited if they are not resolved. The idea of shame and negativity around leaving vulnerabilities open was shared by both IT and cybersecurity professionals. Even based on the researcher's own professional experience and the amount of negativity of cybersecurity breaches in the news, typically an organization is looking for someone to blame. This is another topic that could use further exploration, it is possible that vulnerability management is solely done to avoid confrontation, instead of the explicit desire to secure an organization.

    Vulnerability management is a large concept, it includes secure configuration, vulnerability remediation, network security and defense in depth. Participants also brought up *defense in depth* as a separate code but ultimately it relates back to vulnerability management. Participants expressed concern for *technology* and *modernization* related to vulnerability remediation efforts. Another code identified in relation to vulnerability management was *troubleshooting*, this came from the IT participants. And based on the

researchers own experience in IT operations groups, troubleshooting technology after applying patches or secure configurations is vital to users, functionality, and operations of systems.

**Vulnerability Chaining**

One of the most unanticipated responses from participants was the lack of awareness of *vulnerability chaining* by IT and cybersecurity professionals. Since *vulnerability chaining* is still a relatively unresearched term with cybersecurity programs, this leads the researcher to believe an awareness of this term may be required before *vulnerability chaining blindness* can be further explored. It may be necessary for the researcher to determine why IT and cybersecurity professionals are not aware of *vulnerability chaining*. One participant in the software development / IT profession was able to clearly articulate a definition, but the other participants were not able to fully articulate a definition.

**Vulnerability Chaining Blindness**

*Vulnerability chaining blindness* was the focal point and the cornerstone of this research. But to lead up to this terminology, the idea that *linkage blindness* and *vulnerability chaining* were novel concepts was of particular interest. Since most of the participants were introduced to these terms now, there were several pauses in responses from participants and even some requested time to think about the terms. The researcher noted this while annotating responses in the focus group and thought this was an essential body language component to discuss. When initially presenting participants with the idea of *vulnerability chaining*, there was almost 60 seconds between the researcher's introduction of the concept and participant responses.

Each participant silently nodded their heads as they considered the new terminology and possible implications to cybersecurity programs. What was particularly interesting was how the participants responses, they gave careful and thoughtful feedback on the terminology. Each participant felt that it may have some applicability to cybersecurity programs, even if they did not think that they would use it every day. These responses prove that the phenomenon exists, and that the new terminology would be impactful and potentially change the way vulnerability remediation efforts are conducted.

**IT and Cybersecurity Concurrence**

The researcher found through asking the final two questions in the focus group, that IT and cybersecurity professionals did not necessarily agree on *vulnerability chaining blindness*. While both groups felt that the concept was vital and the terminology should be used, it was unanimous that this should be done on the cybersecurity side. There was concern expressed from the IT operations side that *vulnerability chaining blindness* would not be as applicable to them because of their focus on users and systems owners. The IT professionals agreed that functionality and concerns over technology were more valuable than focusing on how low and medium vulnerabilities could impact their systems.

Another interesting note was that IT and cybersecurity professionals wanted to focus on the *big picture* and have a *holistic view of security*. This was a running theme throughout the focus group and was introduced when *vulnerability chaining* was described to participants. This data suggests that IT and cybersecurity professionals do agree that security should be viewed from a risk management perspective and not getting too far into the weeds. However, *defense in depth* was also brough up a few times, so the *complexity of cybersecurity programs* is an important discovery. Both IT and cybersecurity professionals must ingest so much different types of information to properly secure their systems. It is possible there is more research to be done on understanding just how complex cybersecurity programs are, and potentially why *vulnerability chaining blindness* is not as large of a concern now.

**KEY POINTS:**

1. Vulnerability chaining and linkage blindness were new concepts for IT and cybersecurity professionals
2. Vulnerability chaining blindness does describe the phenomenon of IT and cybersecurity professionals' inability to view and remediate multiple vulnerabilities in combination

3. IT and cybersecurity professionals disagree that vulnerability chaining blindness should be incorporated into cybersecurity programs
4. IT professionals were concerned about how this concept may reflect on their reputations in the industry